\documentclass{elsarticle}

\usepackage{subfigure}
\usepackage{tikz}
\usepackage{circuitikz}
\usepackage{amsmath}
\usepackage{url}
\usepackage{lineno}

\begin{document}

\title{Electric Field Measurement by Edge Transient Current Technique on Silicon Low Gain Avalanche Detector}

\author[label1,label2,label3,label4]{Chenxi Fu}
\author[label4]{Haobo Wang}

\author[label2]{Ryuta Kiuchi}

\author[label5]{Jianing Lin}

\author[label2]{Xin Shi}
\author[label2]{Tao Yang}

\author[label6]{Xiaoshen Kang}
\author[label4]{Weimin Song}
\author[label2]{Congcong Wang}
\author[label1,label7]{Weiwei Xu}

\author[label2]{Zijun Xu}

\author[label1,label7]{Suyu Xiao\corref{cor1}}
\ead{suyu.xiao@iat.cn}

\affiliation[label1]{Shandong Institute of Advanced Technology, 1501 Panlong Road, Jinan 250100, China}
\affiliation[label2]{Institute of High Energy Physics, Chinese Academy of Sciences, 19B Yuquan Road, Shijingshan District, Beijing 100049, China}
\affiliation[label3]{School of Physics Sciences, Institute of Physics, University of Chinese Academy of Sciences, Chinese Academy of Sciences, 3rd South Zhongguancun Street, Haidian District, Beijing 100190, China}
\affiliation[label4]{Jilin University, 2699 Qianjin Street, Changchun 130216, China}

\affiliation[label5]{Beijing National Laboratory for Condensed Matter Physics, Institute of Physics, University of Chinese Academy of Sciences, Chinese Academy of Sciences, 3rd South Zhongguancun Street, Haidian District, Beijing 100190, China}

\affiliation[label6]{Liaoning University, 66 Chongshanzhong Road, Huanggu District, Shenyang 110036, China}
\affiliation[label7]{Shandong University, 27 Shanda Nanlu, Jinan 250100, China}

\cortext[cor1]{Corresponding Author}

\begin{abstract}
A novel methodology, named the diffusion profile method, is proposed in this research to measure the electric field of a low gain avalanche detector (LGAD).
The proposed methodology utilizes the maximum of the time derivative of the edge transient current technique (edge-TCT) test waveform to quantify the dispersion of the light-induced carriers.
This method introduces the estimation of the elongation of the carrier cluster caused by diffusion and the divergence of the electric field force during its drift along the detector.
The effectiveness of the diffusion profile method is demonstrated through the analysis of both simulated and measured edge-TCT waveforms.
Experimental data was collected from a laser scan performed on an LGAD detector along its thickness direction.
A simulation procedure has been developed in RASER (RAdiation SEmiconductoR) to generate signals from LGAD.
An assumption of immediate one-step carrier multiplication is introduced to simplify the avalanche process.
Simulation results were compared with transient current data at the waveform level and showed a favorable match.
Both simulation and experimental results have shown that the diffusion profile method could be applied to certain edge-TCT facilities as an alternative of electric field measurement.
\end{abstract}

\begin{keyword}
	low gain avalanche detector \sep edge transient current technique \sep diffusion profile
\end{keyword}

\maketitle

\section{Introduction}

High-energy experiments, such as the High Luminosity Large Hadron Collider (HL-LHC) \cite{HL-LHC,ATLAS_upgrade,CMS_upgrade}, undergo periodic updates.
In order to meet the requirements of high luminosity facilities \cite{MTD_upgrade_CMS,ITk_upgrade}, there has been a significant increase in the need to enhance and improve particle trackers, particularly in terms of time performance.
Low Gain Avalanche Detectors (LGAD) \cite{LGAD_origin} have gained significant attention as potential candidates for various applications due to their exceptional timing capabilities. It has been reported that Silicon LGAD are able to achieve timing resolutions as low as 20 ps \cite{HPK_time_res, HPK_timing_perf}.
In order to ensure the optimal performance of LGAD, it is crucial to carry out comprehensive testing and measurements before their implementation \cite{HPK_timing_perf, BNL}.

The Transient Current Technique (TCT) is a method used to generate a signal on particle detectors for the purpose of investigating their properties. This technique has been employed for various purposes, such as determining the effective doping level of detectors \cite{TCT_original} and evaluating the gain rate of LGAD \cite{TCT_gain}.
Additionally, the determination of the built-in electric field in p-i-n planar and strip detectors can be achieved by modifying the direction of the injected laser using the Edge Transient Current Technique (Edge-TCT) \cite{Edge-TCT}, by analyzing the carrier velocity profile \cite{vel_prof}.
Nevertheless, the direct application of this method to LGAD is not feasible due to specific limitations. Instead, this study introduces a novel approach called the diffusion profile method.
TCT generates non-equilibrium carrier clusters within the detector using a laser.
Before the carrier avalanche, the distribution of clusters undergoes modifications during the drift process. This phenomenon is the result of diffusion and the divergence of the electric field force.
A flattened carrier distribution results in a signal with a weaker peak, indicating a lower intensity of the electric field or a higher degree of field variation.

The RAdiation SEmiconductoR (RASER) is a Python package designed to simulate physics processes in semiconductor detectors \cite{raser}. The software provides extensibility and customization, distinguishing it from other detector simulators such as Allpix Squared \cite{ap2}, Weightfield2 \cite{wf2}, and TRACS \cite{TRACS_Paper}, which are coded in C++.
The package has shown strong agreement with experimental data concerning the temporal performance of non-irradiated p-i-n diodes \cite{RASER_pin}. Furthermore, previous studies utilizing this package have shown successful predictions for 3D detectors \cite{RASER_3D}.
The aim of employing RASER simulation in this research is to replicate the essential steps of the edge-TCT process in p-i-n and LGAD in order to verify the effectiveness of the diffusion profile method and establish a foundation for future research.
This study represents the initial attempt to model an LGAD as an active device capable of responding to laser beam input. A simplified model for carrier multiplication in an LGAD has been successfully developed in this study. Our approach is based on the underlying assumption that a single carrier entering the gain layer generates $M$ carrier pairs at a specific space-time coordinate. Here, the variable $M$ represents the rate of gain.

\section{Facilities}

\subsection{Detector under test}

\begin{figure*}[htbp]
    \centering
    \subfigure[I-V relation of p-i-n and LGAD.]{
        \centering
        \includegraphics[width=0.48\linewidth]{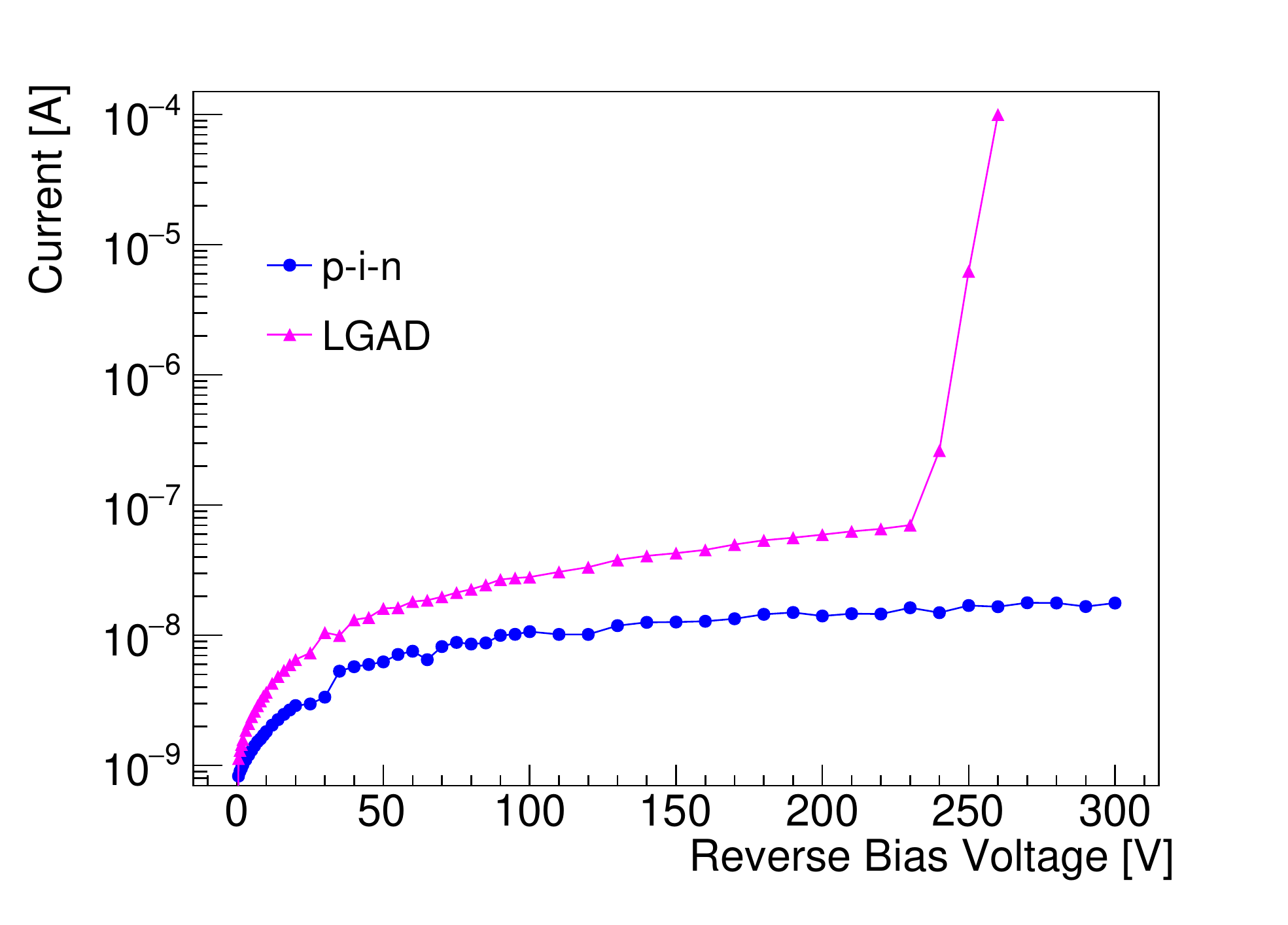}}
    \subfigure[C-V relation of p-i-n and LGAD.]{
        \centering
        \includegraphics[width=0.48\linewidth]{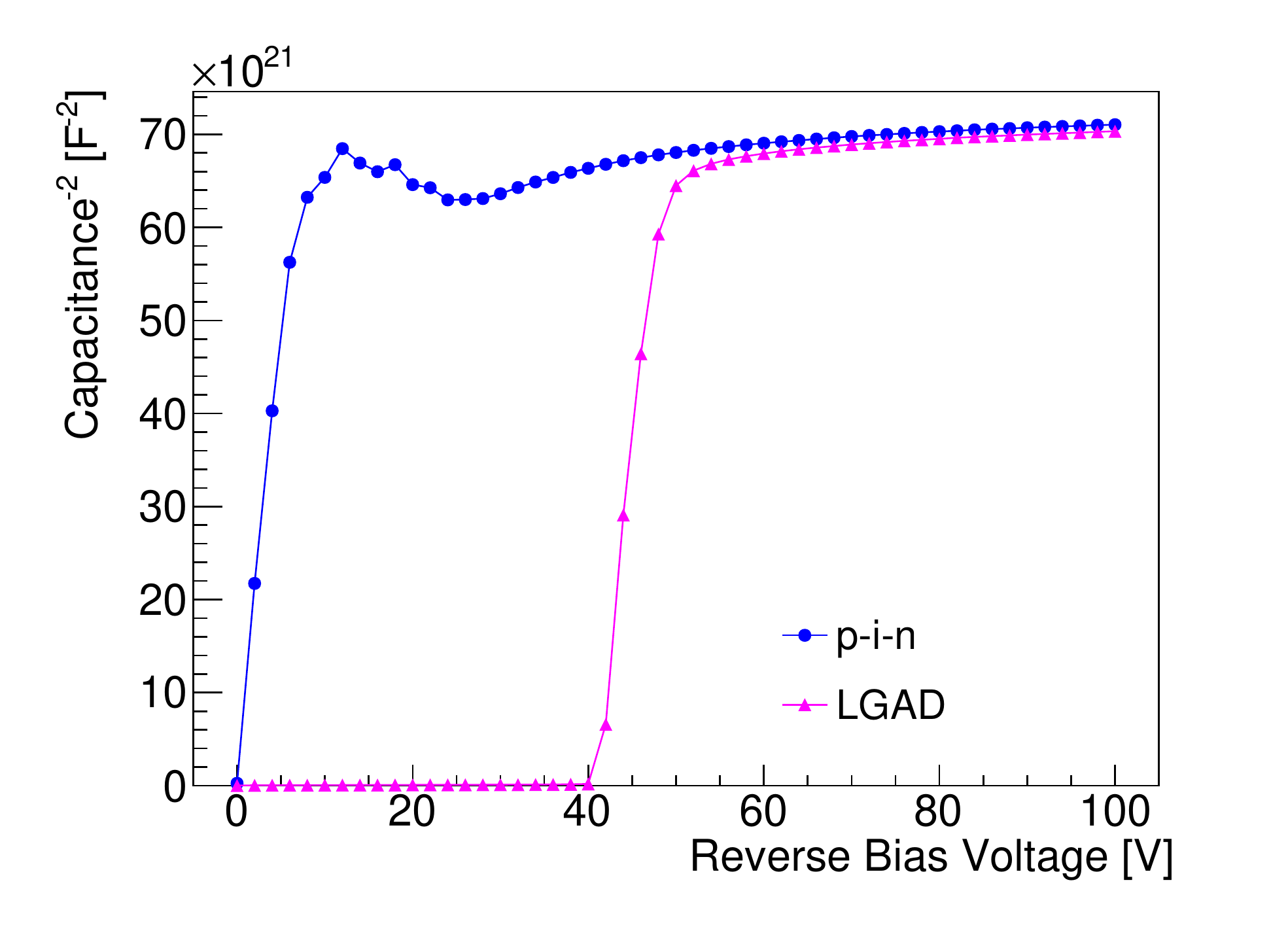}}
    \label{fig:ivcv}
    \caption{I-V and C-V relations of p-i-n and LGAD under test.}
\end{figure*}

The LGAD being tested, along with the p-i-n detector used for comparison, was manufactured by Hamamatsu Photonics K.K. (HPK) \cite{HPK}, specified as HPK Type 3.1-50. The p-type bulk of the detectors has an active thickness of $50\ \rm \mu m$ and a size of $1300 \times 1300\ \rm \mu m^2$. Guard rings are distributed outside the active area.
The I-V and C-V characteristics were measured under reverse bias, as shown in Figure \ref{fig:ivcv}. The experimental setup consisted of a Keithley 2410 power supply, a Keithley 6487 current meter, and an Agilent E4980A LCR meter operating at a frequency of $10\rm kHz$.

The abrupt junction model has been used to estimate the distribution of doping.
Both the p-i-n and LGAD structures exhibit a bulk doping concentration of $1\times 10^{12}\ \rm cm^{-3}$. Additionally, the LGAD structure incorporates a deep-doped region with a concentration of $1.913 \times 10^{18}\ \rm cm^{-3}$, which is estimated from C-V data and the gain rate. The deep-doped area is specifically modeled within the LGAD structure, covering a range of $1 \sim 2\ \rm \mu m$.

\subsection{Experiment Setup}

\begin{figure}[htbp]
	\includegraphics[width=0.6\linewidth]{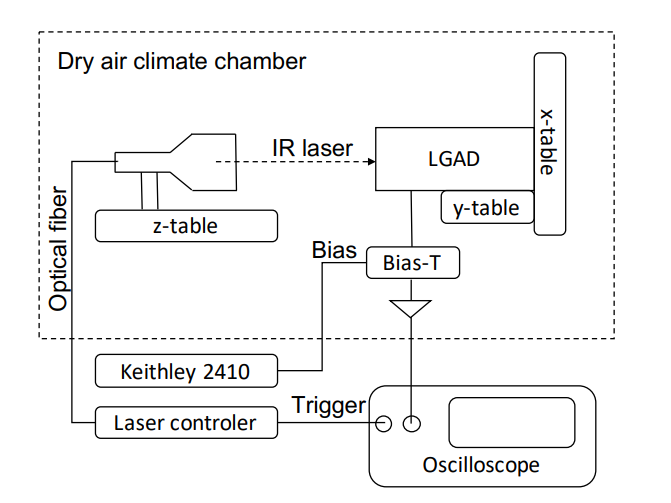}
	\centering
	\caption{Schematic view of the edge-TCT setup at CERN RD50 collaboration \cite{RD50}. Experiment data recorded in 2019.}
	\label{fig:eTCT set}
\end{figure}

The experimental setup for the edge-TCT scan is illustrated in Figure \ref{fig:eTCT set}.
An infrared laser with a wavelength of 1064 nm is utilized, which has an attenuation length of approximately 1 mm in silicon. The given length is sufficient for the laser beam to effectively penetrate the detector. The beam waist of the pulse has a full width at half maximum (FWHM) measurement of 8 $\rm \mu$m, and the pulse duration is 350 ps.
Denoting the direction of the laser beam as the $x$-axis, the direction along the edge as the $y$-axis, and the direction of the electric field inside the detector as the $z$-axis serves as a convenient coordinate system for analysis. The value of z is initialized to zero at the top of the detector.
The laser beam should be positioned perpendicular to the edge surface of the detector.
Three portable tables have been arranged to facilitate the adjustment of the laser's focal point, enabling a comprehensive z-scan across the entire thickness of the detector (0 to 50 $\rm\mu$m) with a step size of 1 $\rm\mu$m. Before capturing the laser-induced waveform, the laser beam is precisely directed towards the edge of the active area of the detector. This is achieved by adjusting the focus of the laser in a direction perpendicular to the laser beam at different positions along the z-axis. This ensures that the generated electron-hole pairs are within a restricted range near the designated value of $z$. The resulting variation in waveform amplitude is then observed to determine the minimum width of the laser beam.
A Keithley 2410 source is used as a high-voltage power supply. To enhance the signal, a CIVIDEC \cite{cividec} wideband amplifier with a bandwidth of 2 GHz, a gain of 40 dB, and an integrated bias-Tee is utilized.
After shaping, the signal was converted into a digital format and recorded using a Keysight DSO9254A oscilloscope \cite{oscil}, which has a bandwidth of 20 GHz and a sample rate of 20 GSa/s.

In the following TCT tests, the devices were measured at room temperature. Except for gain measurement, the devices were biased at 200 V from the backside.

\section{Modeling}
\label{3}

The modeling of the edge-TCT process can be categorized into four main components: field solution, generation of laser-induced  non-equilibrium carriers, collection of carrier drift signals, and filtering of the final waveform.
For LGAD, an additional step of carrier multiplication is implemented.

\subsection{Field profile}

The electric potential $U$ of a detector can be calculated by solving the Poisson equation

\begin{equation}\label{eq:U}
\nabla^2 U(\vec x) = -eN_{eff}(\vec x)/\varepsilon,
\end{equation}
in which $\vec x = (x,y,z)$ represents the vector coordinates, $e$ denotes the unit charge, $\varepsilon$ denotes the dielectric constant of the detector material, and $N_{eff}$ denotes the effective doping concentration.
The depletion ansatz is used to simplify the calculation.
In addition, it is assumed that the depletion area of the heavily-doped Ohmic contact layers is thin enough to be considered negligible.

The calculation of the weighting potential for signal generation is performed in a similar manner, using the following equation

\begin{equation}\label{eq:U_w}
\nabla^2 U_w(\vec x) = 0.
\end{equation}
At the readout electrode, $U_w$ is set to 1, while on all other electrodes, $U_w$ is set to 0.

Within regions of high electric fields, carriers undergo acceleration and gain enough energy to ionize electrons in the valence band. The multiplication behavior of carriers is determined by the ionization rate, which is defined as the number of electron-hole pairs generated by a drifting carrier as it travels a unit distance.
The ionization rate for electrons ($\alpha_n$) and holes ($\alpha_p$), which is determined by the temperature $T$ and intensity $E$, can be expressed as \cite{VANOVERSTRAETEN1970583}

\begin{equation}\label{eq:alpha}
\alpha_{p,n}(T,E) = \gamma(T) a_{p,n}\exp (-\frac{\gamma(T)b_{p,n}}{E}),
\end{equation}
where $a_{p,n}$ and $b_{p,n}$ are fixed parameters, and
$\gamma(T)=\tanh (\hbar \omega/2kT_0)/\tanh \left[\hbar \omega /2kT (T/T_0)\right]$
is a correction term for phonon-electron interaction, with $T_0 = 300\rm \ K$ and $\hbar \omega$ denoting the average phonon energy within the material \cite{seeger_2011}.

\subsection{Laser and carrier generation}
\label{3.2}

The rate at which light-induced carriers are generated is proportional to the intensity of the laser beam. For a typical Gaussian laser beam in the $x$ direction, the light intensity distribution is \cite{laser}

\begin{equation}
I(r,x,t)=\frac{E_p}{\Delta t}\frac{4\sqrt{\ln 2}}{\pi ^{3/2}w^2(x)}\exp{\frac{-2r^2}{w^2(x)}}\exp{\frac{-4t^2\ln2}{(\Delta t)^2}}
\label{laser_intensity}
\end{equation}
where $r^2 = y^2+z^2$, $E_p$ and $\Delta t$ are the single pulse energy and the pulse duration of the laser, and $w_0=\sqrt{\lambda/\pi NA},w(x)=w_0 \sqrt{1+(\lambda x/\pi w_0^2 n)^2}$ are the beam widths at $0$ and $x$ (with the origin at the beam focus), determined by the numerical aperture ($NA$) of the focusing lens and the laser wavelength ($\lambda$). The spatial FWHM of the laser beam is related to $w(z)$ by $\mathrm{FWHM}_{spatial} = w(z) \sqrt{2\ln{2}}$.

Considering refraction and absorption, the intensity function will change its form into

\begin{equation}
I'(r,x,t)=I_0(r,x,t)\exp\left[ -\alpha(\lambda)x\right].
\label{eq:light_absorb}
\end{equation}
The photons emitted by the laser beam are converted into excitation energy for the carriers. During a small time interval $dt$, the density of generated carriers can be calculated by

\begin{equation}
\mathrm{d}N_0^{e-h\ pairs}=\alpha I'\mathrm{d}t/h\nu,
\end{equation}
where $h\nu$ represents the energy of a photon.

\subsection{Carrier drift and multiplication}

Once the carriers are excited, they will drift through the bulk material, thereby generating an electric current. According to the mobility formula \cite{Baliga}

\begin{equation}
v_{p,n} = \mu_{p,n}(T,N_{eff},E) E,
\label{eq:velocity}
\end{equation}
the velocity of a carrier at a given position can be mathematically represented as a function of the effective doping concentration $N_{eff}$ and the electric field intensity $E$.
In this study, the Reggiani model \cite{Reggiani} is employed. 
Aside from oriented drifting, a Langevin term is considered to simulate the random motion of the carriers by adding a Gaussian displacement $\mathrm{d} \vec x_{Langevin}$, the probability distribution function of which is

\begin{equation}
	f(\mathrm{d} \vec x_{Langevin} ) = \frac{1}{(2\pi D\mathrm{d} t)^{3/2}}\exp{\left(-\frac{|\mathrm{d}\vec x_{Langevin}|^2}{2D\mathrm{d} t}\right)}
\end{equation}
where

\begin{equation}
	D = k_B T\mu /q
\end{equation}
is the Einstein diffusion coefficient. Here, $k_B$, $T$, and $q$ refer to the Boltzmann constant, temperature, and elementary charge, respectively. 

According to Shockley-Ramo's theorem, a signal $i_q(t)$ will be generated at the reading electrode during the drift of a charge $q$ by

\begin{equation}
i_q(t) = q \vec v_q(t) \cdot \nabla U_w[\vec x_q(t)].
\label{eq:induced_current}
\end{equation}
The complete signal waveform can be acquired by combining the signals from all generated carriers.
Additionally, in LGAD, a carrier in the gain layer induces carrier multiplication, which leads to signal amplification.

\begin{figure}[htbp]
    \centering
    \includegraphics[width=0.98\linewidth]{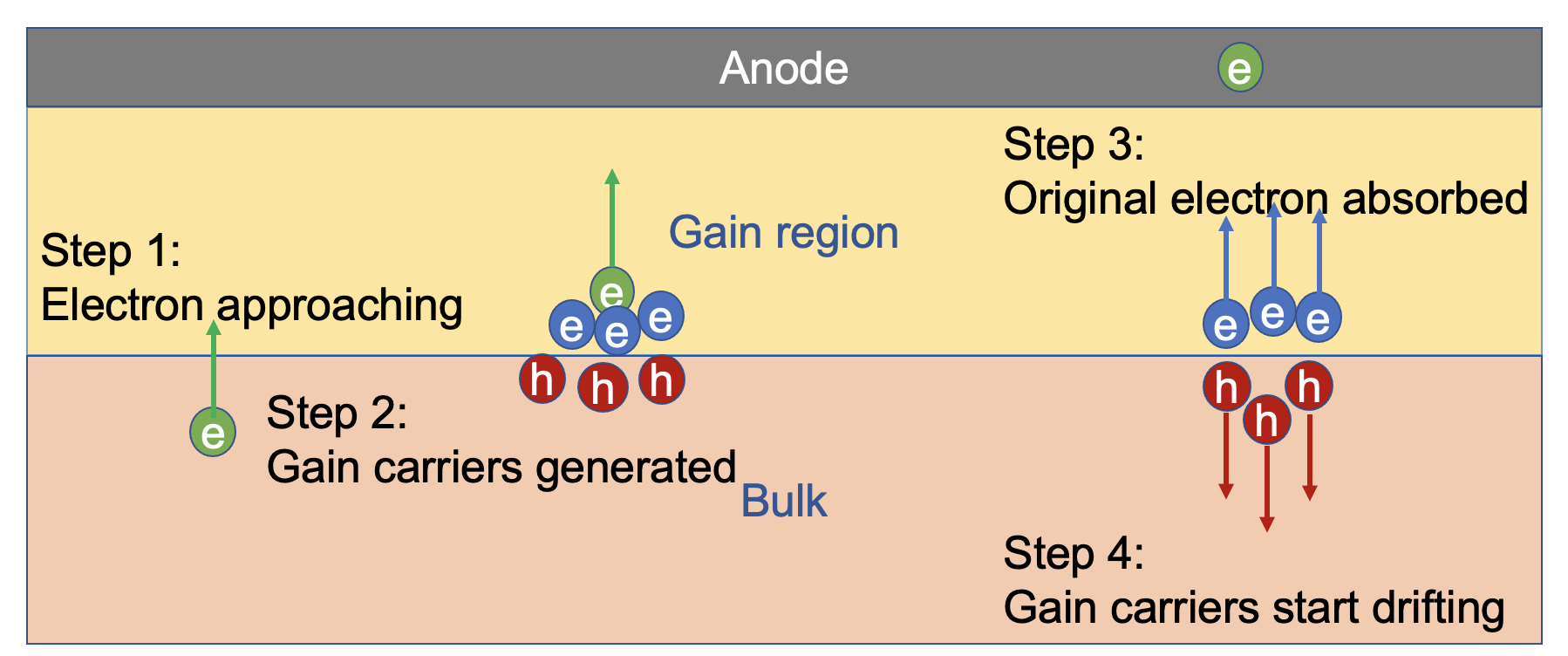}
    \caption{Sketch of simplified carrier multiplication process in RASER, limited at the boundary between gain layer and the bulk. This picture is not to scale.}
    \label{fig:gain_sketch}
\end{figure}

Instead of simulating the avalanche step by step, as illustrated in Figure \ref{fig:gain_sketch}, it is assumed that the entry of a single carrier into the gain layer will result in the generation of $M$ pairs of carriers without ionizing capability at the bottom of the gain layer, where

\begin{equation}\label{eq:M_gain}
M=\frac{\exp \left[\int_{0}^{d_{gain}}\left(\alpha_{\mathrm{n}}-\alpha_{\mathrm{p}}\right) \mathrm{d} x\right]}{1-\int_{0}^{d_{gain}} \alpha_{\mathrm{p}} \exp \left[\int_{0}^{x}\left(\alpha_{\mathrm{n}}-\alpha_{\mathrm{p}}\right) \mathrm{d} x\right] \mathrm{d} x}
\end{equation}
is the gain factor of the device \cite{Baliga}. As the values of $\alpha_n$ and $\alpha_p$ are sufficiently small outside the designated gain layer, it is advantageous to only integrate within the gain layer, which has a width of $d_{gain}$. The integral diverges when the device breaks down. For the LGAD under test, the value of $M$ is determined to be 23.4.

Due to the double-exponential relationship between the field intensity $E$ and the parameter $M$ (see Eq. \ref{eq:alpha} and \ref{eq:M_gain}), even a minor modification in the doping profile can result in a substantial change in the gain rate.
This statement elucidates the challenges associated with the fabrication process of LGAD. However, in simulation, it is possible to make a numerical adjustment to the doping profile (by $10^{12} \rm\ cm^{-3}$) in order to align the gain rate value with experimental results.

\subsection{Signal collection}

\begin{figure}[htbp]
    \centering
    \includegraphics[width=0.98\linewidth]{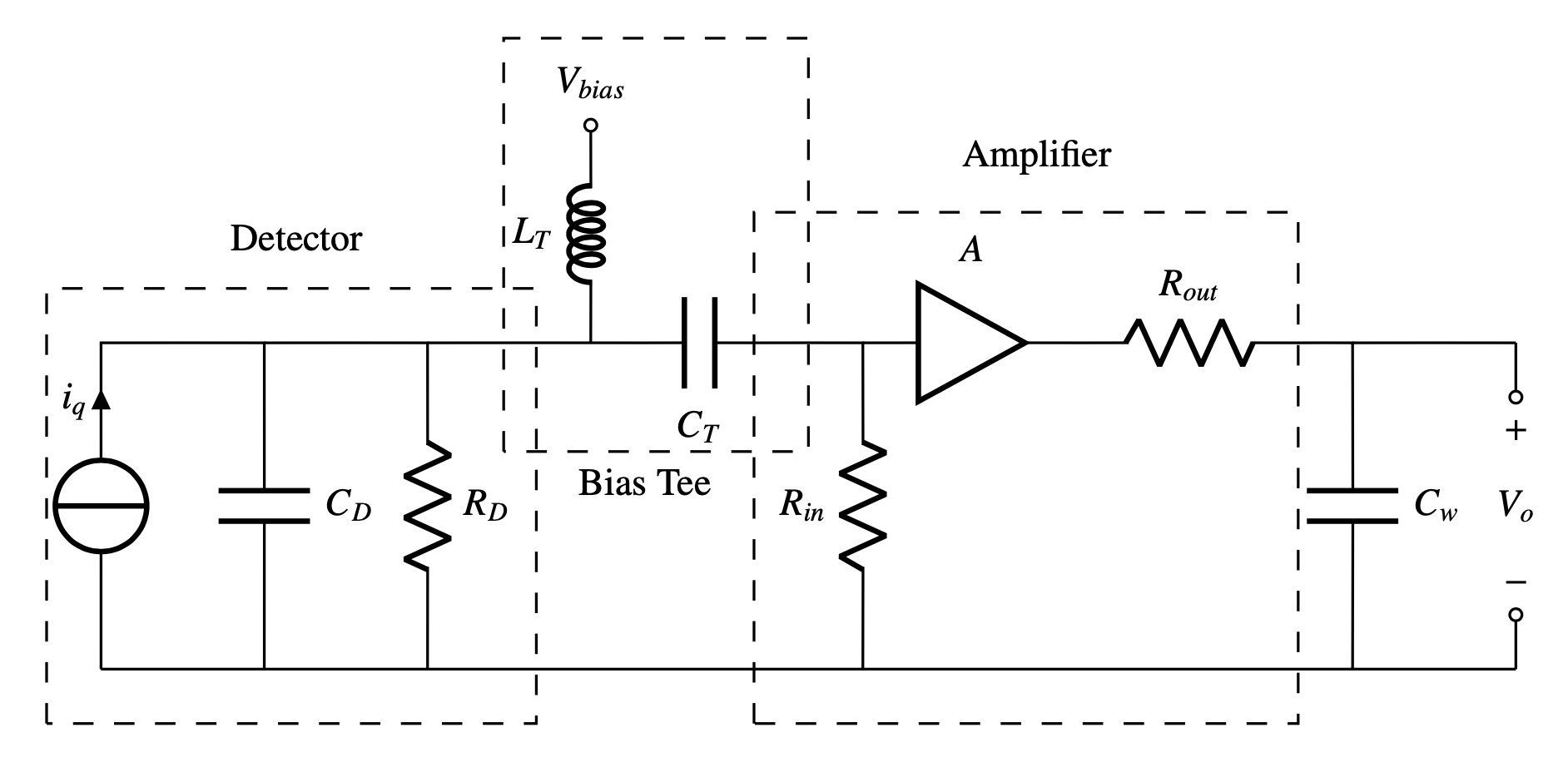}
    \caption{Equivalent circuit of Figure \ref{fig:eTCT set}.}
    \label{fig:circuit}
\end{figure}

Amplification is needed for the original induced currents to be detected, stored, and analyzed.
The electronic system can be modeled as shown in Figure \ref{fig:circuit}, which incorporates the signal effects of various devices. These effects include the capacitance $C_D$ and the AC resistance $R_D$ of the detector. Additionally, the inductance $L_T$ and capacitance $C_T$ of the bias-tee contribute to these effects. Furthermore, the gain ($A$) of the amplifier, the input resistance ($R_{in}$), the output resistance ($R_{out}$), and the capacitance of the cable ($C_w$) also contribute to these effects.

\tikzset{global scale/.style={
    scale=#1,
    every node/.append style={scale=#1}
  }
}

In the frequency domain, if $1/R_D$ and $1/L_T$ are small enough to be neglected, the transfer function of the aforementioned devices can be expressed as (letting $s = j \omega$)

\begin{equation}
H(s) = \frac{R_{in}C_T}{C_T+C_D+R_{in}sC_TC_D}\frac{A}{sC_wR_{out}+1},
\end{equation}
and by using Laplace transformation, the amplification and shaping of the original signal can be expressed through convolution with

\begin{equation}
h(t)=A^* \frac{\exp(-t/\tau_{RC_D}) - \exp(-t/\tau_{RC_w})}{\tau_{RC_D}-\tau_{RC_w}} ,
\end{equation}
where $\tau_{RC_D} = R_{in}C_T C_D/(C_T+C_D)$ and $\tau_{RC_w} = R_{out}C_w$ are the time constants of the circuit. Additionally, $A^* =A\tau_{RC_D}/C_D$ is the total amplification rate, measured in units of $\Omega$, that converts the original current signal into voltage.

\section{Waveform Simulation}

This section aims to evaluate the results of the edge-TCT z-scan in the direction of the device thickness.
A global normalization factor is applied to both p-i-n and LGAD. For LGAD, the laser intensity is 0.58 times stronger than that of p-i-n in order to obtain a more distinct waveform.

\begin{figure*}[htbp]
    \centering
    \subfigure[Edge-TCT waveform of p-i-n.]{
        \centering
        \includegraphics[width=0.47\linewidth]{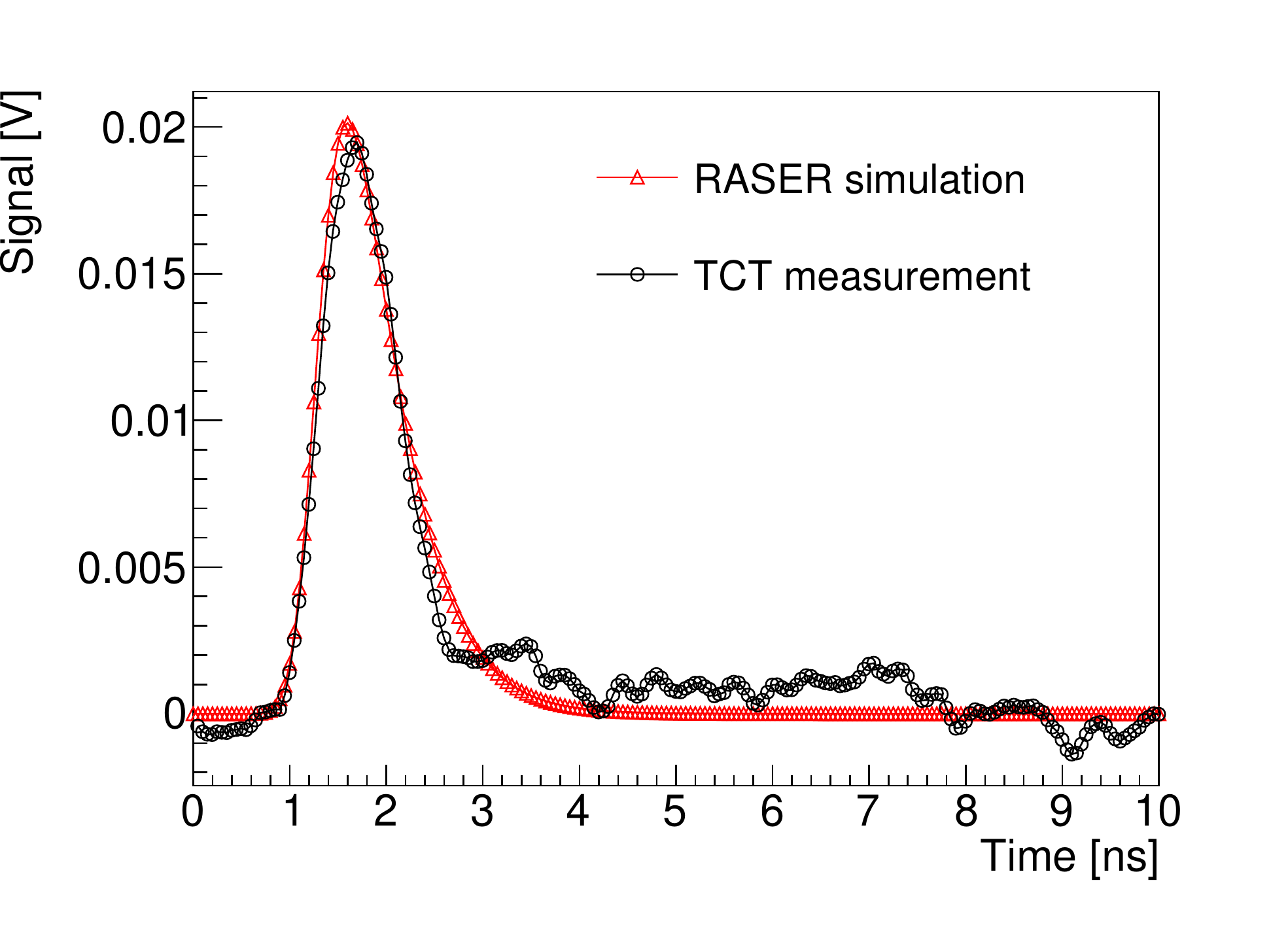}
    }
    \subfigure[Edge-TCT waveform of LGAD.]{
        \centering
        \includegraphics[width=0.47\linewidth]{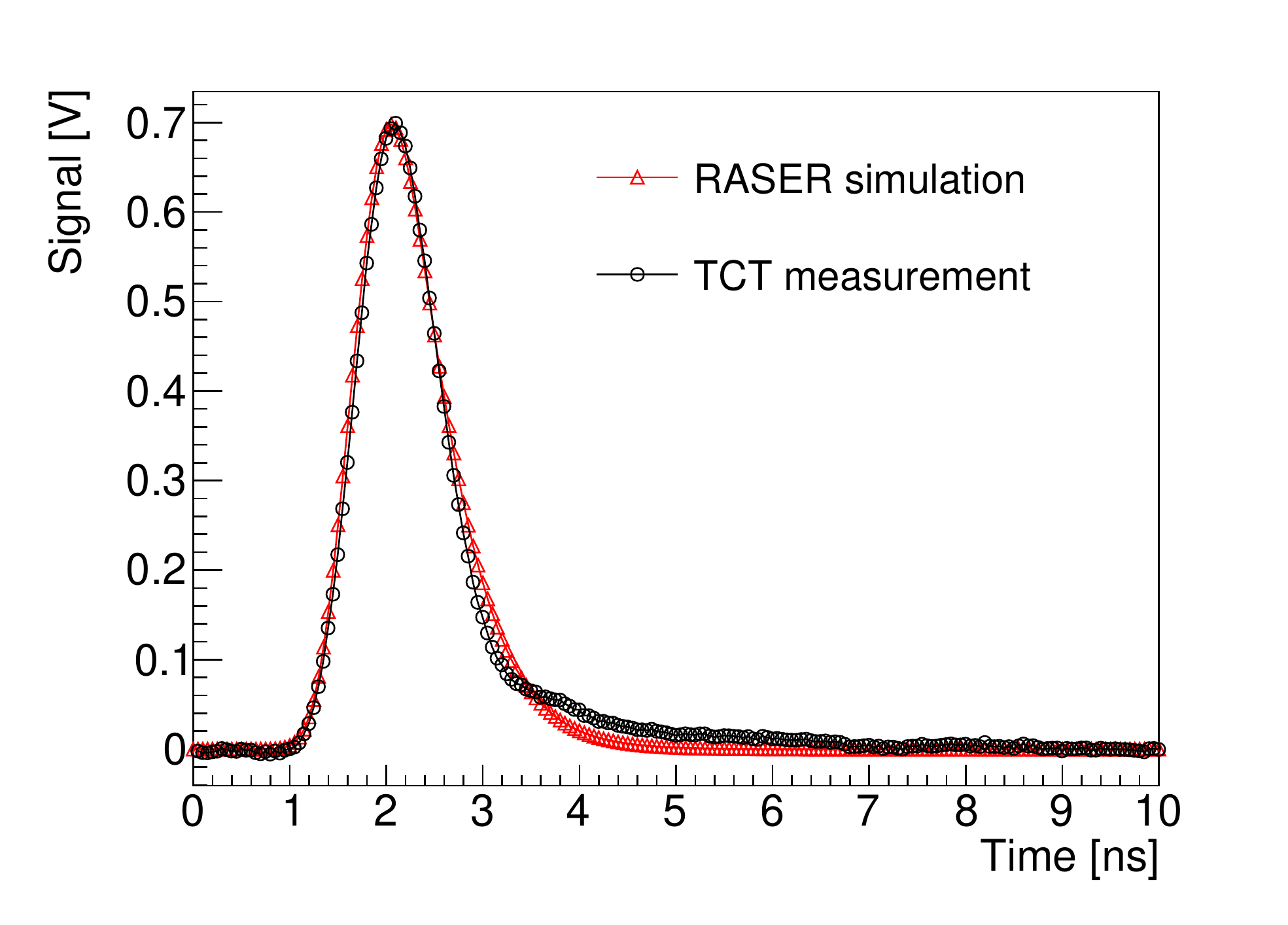}
        \label{fig:waveform:lgad}
    }
    \caption{Edge-TCT waveform of both detectors, laser focus at $z$ direction of $25\ \rm \mu m$, bias voltage 200 V.}
\label{fig:waveform}
\end{figure*}

In Figure \ref{fig:waveform}, a good match of waveforms is exhibited when comparing the simulation and experimental results of p-i-n and LGAD.
But to ensure reliable edge-TCT scan simulation, the waveform quantities, such as amplitude, rise time, and charge collection efficiency, should have a good agreement with experimental results at every point.

\begin{figure*}
    \centering
    \subfigure[Amplitude-z relation.]{
        \centering
        \includegraphics[width=0.47\linewidth]{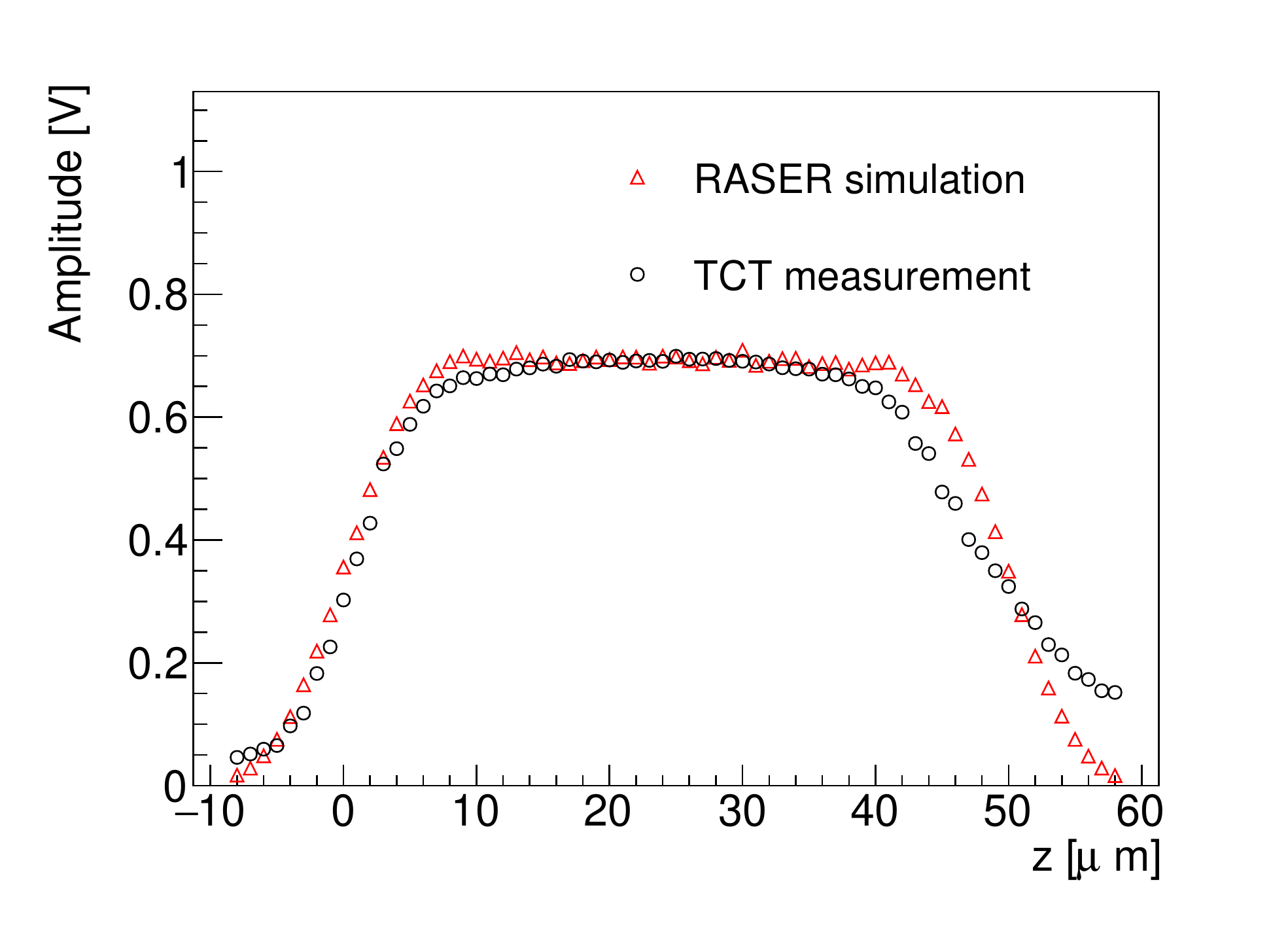}
        \label{fig:amplitude}
    }
    \subfigure[Rise time-z relation.]{
        \centering
        \includegraphics[width=0.47\linewidth]{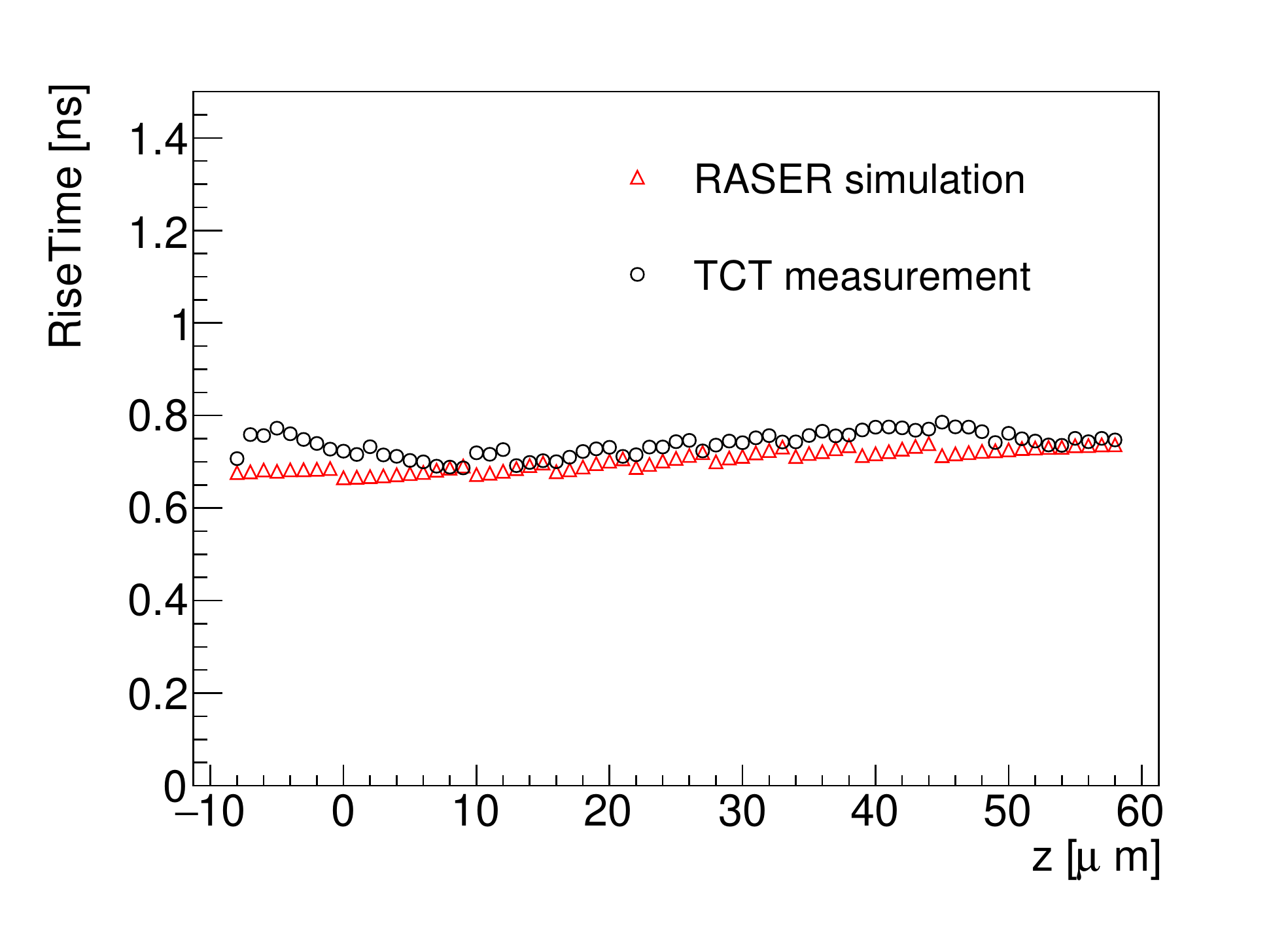}
        \label{fig:risetime}
    }
    \quad
    \subfigure[Charge collection-z relation.]{
        \centering
        \includegraphics[width=0.47\linewidth]{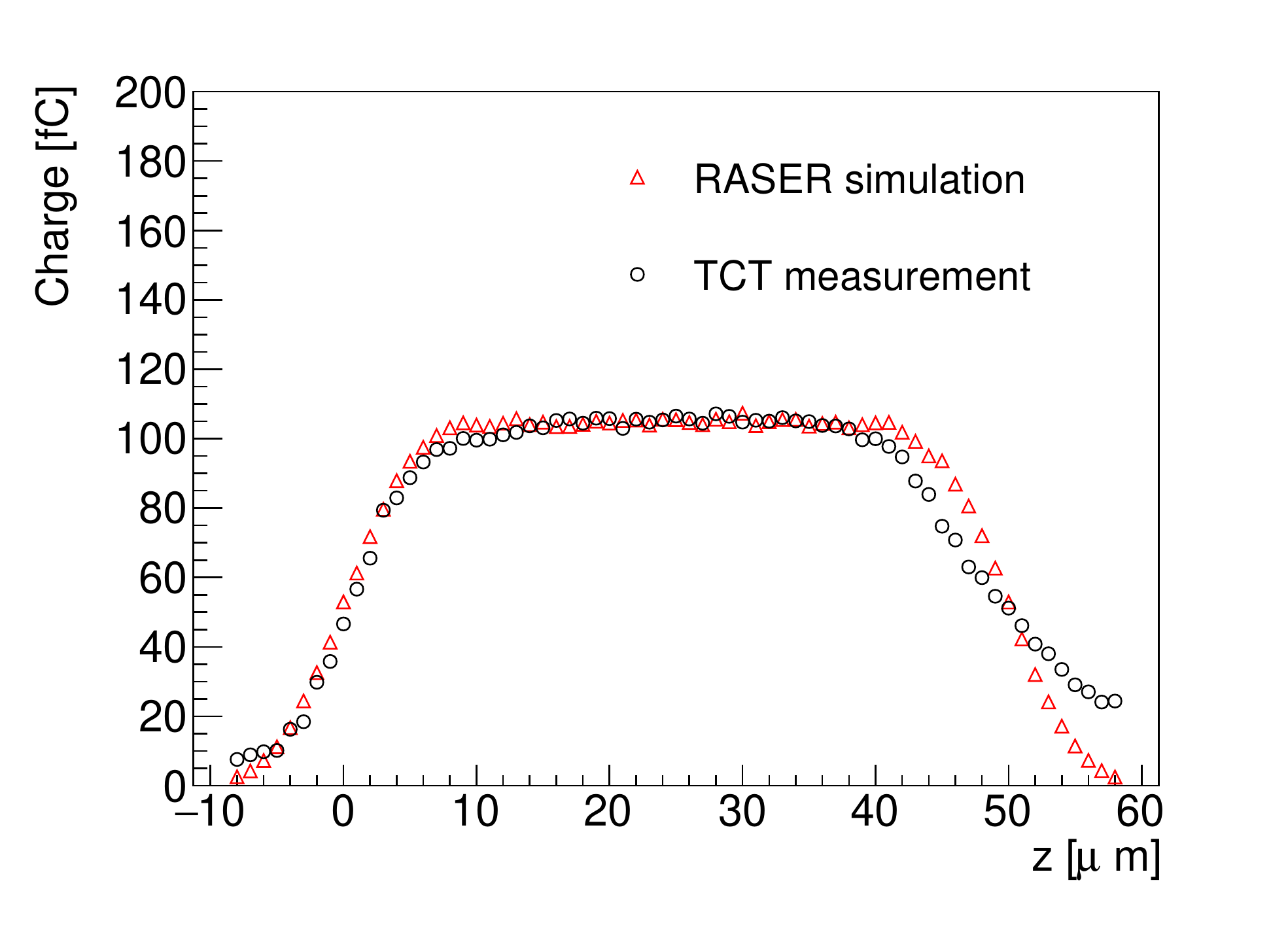}
        \label{fig:charge}
    }
    \subfigure[Charge collection-Bias voltage relation.]{
        \centering
        \includegraphics[width=0.47\linewidth]{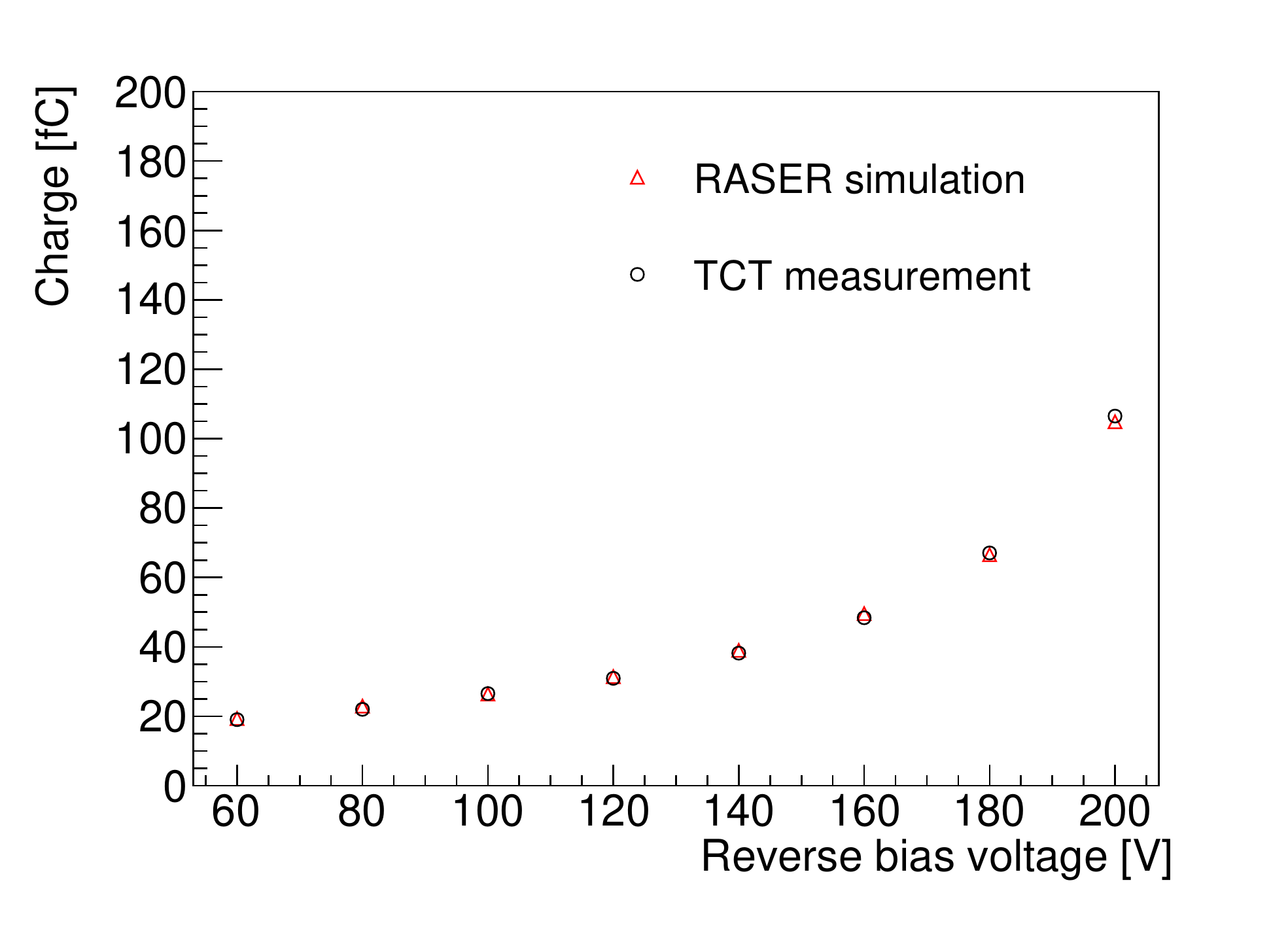}
        \label{fig:gain}
    }
    \caption{Edge-TCT scan results of LGAD, top at $z=0$, bottom at $z=50\ \rm \mu m$. }
\end{figure*}

\subsection {Amplitude}

\label{section:4.1}

Figure \ref{fig:amplitude} illustrates the relationship between the amplitude ($A$) of the waveform and the position of the laser focus ($z$). The simulation results demonstrate an overestimation in the range of $z\in (40,50)$ and an underestimation in the range of $z\in (50,60)$. This phenomenon may potentially be attributed to the diffusion of dopants inside the substrate, resulting in a non-abrupt junction between the substrate and the bulk. Additionally, the lack of consideration for light-induced carriers within the substrate, which diffuse into the active layer, may have also played a role.

\subsection {Rise time}

The rise time of a signal can be determined by performing linear fitting on the data points within the range of 20\% to 80\% of the signal's amplitude. The relationship between the rise time and the variable $z$ is illustrated in Figure \ref{fig:risetime}.
An uniform result of 0.7 ns is observed and simulated for LGAD, indicating stable signal generation and an excellent signal-to-noise ratio. The presence of jaggedness within the figure can be attributed to the use of discrete data points.

\subsection {Charge collection}

In this work, the charge collection of a waveform is defined as the integral of the current over a specific time period. The result of the charge versus focus $z$ coordinate is shown in Figure \ref{fig:charge}.
As mentioned in Section \ref{section:4.1}, apart from the junction and carrier diffusion at the bottom, the charge collection successfully reflected the deposition of laser energy.

\subsection{Gain measurement}

To measure the gain rate of LGAD, we examined the relationship between charge and reverse bias voltage $V$. Figure \ref{fig:gain}, which was sampled at $z=25\rm \ \mu m$. No charge collection change is observed for p-i-n, thus the charge collection of p-i-n is used as a $gain=1$ contract. Under the given voltage range of $[60, 200] \rm \ V$, the gain rate of LGAD increases from 4 to 26. At 200V, the integral in \ref{eq:M_gain} is 23.4, while the simulation result from the waveform is 24, with a relative error of less than 10\%.

With comparisons of several quantities, an edge-TCT z-scan of silicon LGAD is performed to ensure the accuracy of LGAD simulation. This work solidifies and enhances the method of simplifying multiplication, making it more reliable. It serves as a foundation for future research.

\section{Electric Field Measurement}

\subsection {Velocity profile}

The primary utility of edge-TCT lies in its ability to quantify the electric field within the bulk of the detectors.
The velocity profile method is widely employed, particularly for p-i-n detectors \cite{vel_prof}. For an electron or hole moving within the interior of the detector's active area, its velocity is determined by the field intensity, as described by Equation \ref{eq:velocity}. The signal, which is primarily determined by the drifting velocity (excluding trap capture), can be expressed using Equation \ref{eq:induced_current}. That is, it has the capability to extract field intensity information from the edge-TCT waveform.

Considering a pair of carriers generated at a specific location $z=z_0$, the sum of their velocities at the instant they are generated is

\begin{equation}
\begin{aligned}
v_e(t) + v_h(t) &= \mu_e[E(z_e)] E(z_e) + \mu_h[E(z_h)] E(z_h) \\
&\approx \{\mu_e[E(z_0)]+\mu_h[E(z_0)]\} E(z_0).
\end{aligned}
\end{equation}
If a group of carriers conforms to the assumption that they are produced within a limited region and time period around $t_0$, then the resulting signal they generate should be

\begin{equation}
i_q(z;t\to t_0)=\sum _{q}q \vec v_q\nabla U_w\approx \frac{ne}{d}(\mu_e+\mu_h)E(z)
\label{eq:velocity_proof}
\end{equation}
as $\nabla U_w \approx (1/d) \ \hat z$, and $\vec E \approx E \hat z$ in planar structure detectors.
After amplification and RC shaping, it can be observed that this proportional relationship may not hold strictly true. However, the functional relationship $I(z; t \to t_0) \approx I(E(z))$ should remain valid.

\begin{figure*}[htbp]
    \centering
    \subfigure[Velocity profile of LGAD.]{
        \centering
        \includegraphics[width=0.47\linewidth]{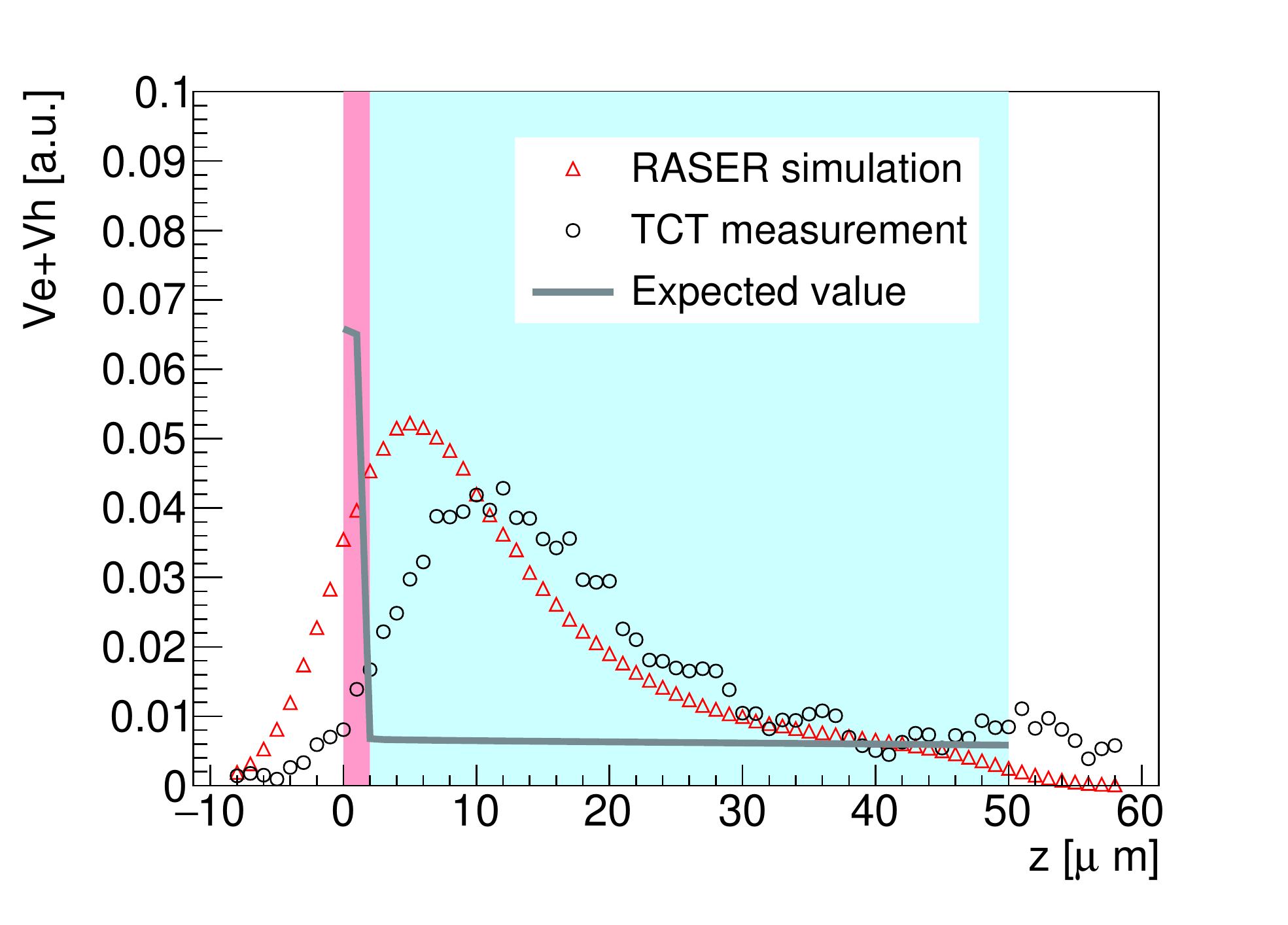}
        \label{fig:velprof}
    }
    \subfigure[Diffusion profile of LGAD. The dashed box represents a preferred range that excludes the effect of the non-Gaussian shape of the carrier cluster.]{
        \centering
        \includegraphics[width=0.47\linewidth]{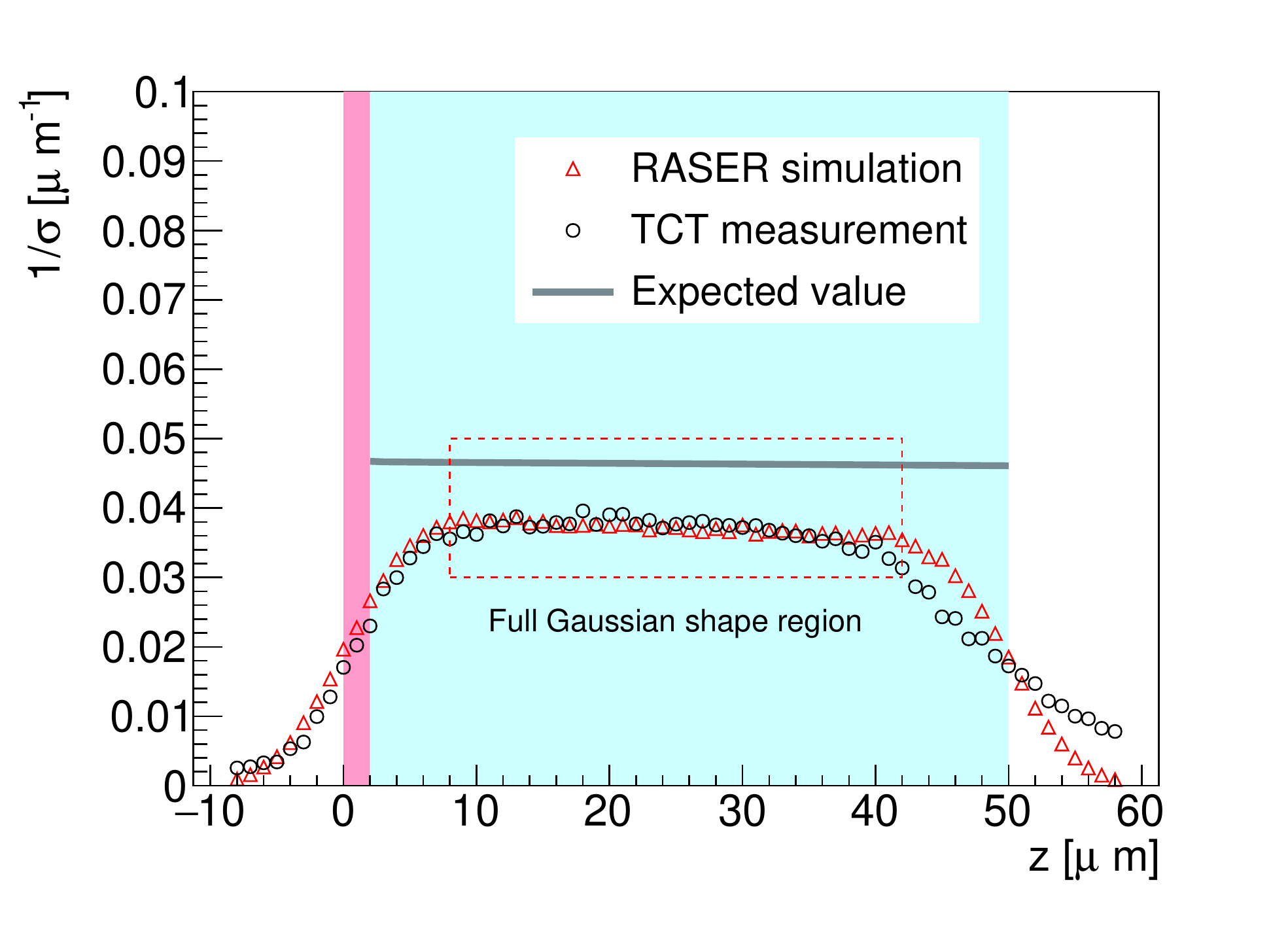}
        \label{fig:diffusion}
    }
    \caption{Two methods for LGAD electric field measurement. The pink region represents the gain layer, while the cyan region represents the bulk.}
\label{fig:field}
\end{figure*}

To counteract the effect of carriers leaving the laser focus and carriers emerging, the velocity profile is chosen as the signal value 0.1 ns after the waveform emerges.
The simulation results of the LGAD, as depicted in Figure \ref{fig:field}, demonstrate a significant peak at $z=5$ for the simulation and $z=10$ for the experiment. Both results show no correlation with the known theoretical LGAD field. The velocity profile method actually loses its effectiveness for LGAD with our facilities. 

This could actually be proven by the analyses provided below. According to Equation \ref{eq:velocity_proof}, the extracted signal, denoted as $i_q$, should exclude the gain signal. However, it is not possible to distinguish between the peak of laser intensity and the peak of the gain signal (see Figure \ref{fig:waveform:lgad}).
To provide a more precise and comprehensive explanation, in systems where the time required for carrier drifting is roughly equivalent to the duration of the laser pulse (both of which are approximately 0.5 ns in our facilities), the "velocity profile" does not provide any information about the coordinates of the focal point. Consequently, it is not suitable for quantifying velocity or field intensity. Additionally, the back-end electronics may also impact the resolution.
For systems of this nature, a specific methodology is required to analyze the internal built-in field, using additional variables. In addition to considering the idealized signal input, an alternative approach involves examining the collective behavior of non-equilibrium carriers induced by light and analyzing their statistical behavior.

\subsection{Diffusion Profile}

In this section, the width of the propagating carrier cluster is being researched.
Ignoring carrier recombination, the distribution of a cluster of light-induced non-equilibrium carriers drifting and diffusing under a uniform electric field is \cite{seeger_2011}

\begin{equation}
N = N_0 \ast \frac{1}{\sqrt{2\pi \sigma_D^2}}\exp{[-\frac{( z-\mu E t)^2}{2 \sigma_D^2}]}
\end{equation}
where $N=N^{e-h\ pairs}(z, t)$, $N_0= N_{0}^{e-h\ pairs}(z, t)$ represents the initial distribution of the carriers as defined in equation \ref{eq:light_absorb}, and $\sigma_D^2 = 2Dt$ if the diffusion coefficient $D$ is constant. The convolution is performed in both the spatial and temporal dimensions.

For an initial spatial-temporal Gaussian distribution with spatial spread $\sigma_0$ and temporal spread $\tau$,
the convolution can be reduced to a variance composition as

\begin{equation}
\begin{aligned}
N(z, t) &= \frac{1}{\sqrt{2\pi \tau^2}}\exp{(-\frac{t^2}{2 \tau^2})}
\\
&\ast \frac{\sum N}{\sqrt{2\pi \sigma^2}}\exp{[-\frac{ (z - z_0 - \mu E t)^2}{2 \sigma^2}]},
\end{aligned}
\label{eq:time_cov}
\end{equation}
where $\sum N$ represents the total number of carriers, $\tau$ denotes the temporal attenuation coefficient of the laser, $\sigma^2 = \sigma_0^2 + 2Dt$, and $\sigma_0$ represents the standard deviation of the initial Gaussian distribution of carriers. The convolution asterisk now only affects the time dimension. 
This analysis is effective for edge-TCT, as the injected Gaussian beam profile behaves like a Gaussian function with $\sigma_0 = w(x)/2$ and $\tau = \Delta t/2\sqrt{2\ln 2}$.

Now consider the elongation of the carrier cluster during drifting. Denote $D = D(z)$ and $E = E(z)$ as the diffusion constant and the field intensity at the center of the carrier cluster, respectively. In an infinitesimal time period $\mathrm{d}t$, the distance between any two carriers would be stretched by the gradient of the field, which can be represented as $\frac{\mathrm{d}(\mu E)}{\mathrm{d}z} \mathrm{d}t$. Additionally, the Einstein coefficient $D$ will change as the carrier cluster center moves, and as a result, the overall distribution will take this form:

\begin{equation}
N(z, t+\mathrm{d}t) = N_0 \exp{\left\{-\frac{[z/(1+\frac{\mathrm{d}(\mu E)}{\mathrm{d} z} \mathrm{d}t)]^2}{2(\sigma^2+ D\mathrm{d}t)}\right\}}.
\end{equation}
Therefore,

\begin{equation}
\frac{\mathrm{d}\sigma^2}{\mathrm{d}t}=\frac{k_BT}{q} \mu  +  2 \sigma^2 \frac{\mathrm{d}(\mu E)}{\mathrm{d}z}.
\end{equation}
Thus, we could integrate (the foot marks $i$ and $e$ represent the initial and end points of the cluster center's path) and obtain

\begin{equation}
\sigma^2
=\frac{(\mu_e E_e)^2}{(\mu_i E_i)^2} \left[\int_{z_i}^{z_e} {\frac{k_BT}{q} \frac{\mathrm{d}z}{E} \frac{(\mu_i E_i)^2}{(\mu E)^2}+\sigma_0^2}\right],
\label{sigma2}
\end{equation}
and if the variation of mobility $\mu$ could be ignored and a uniformly doped bulk is considered (the electric field intensity of which could be calculated by $E(z)=E(z_e)-q N_{eff}(z-z_e)/\varepsilon$), it is revealed that

\begin{equation}
    \sigma^2 = \frac{E_e^2}{E_i^2}\left[\sigma_0^2  +\frac {k_1}{2}\left(1-\frac{ E_i^2}{ E_e^2}\right)\right],
    \label{sigma2_simple}
\end{equation}
where $k_1=(k_B T/q)\cdot(\varepsilon/q N_{eff})= 2.8 \rm \ {\mu m}^2$.

Now consider the multiplication process in LGAD. Notice that from then on, only the carriers leading the avalanche will be considered, as due to the device design, they often have a larger ionization rate.
To simplify calculations, it could be assumed that during the avalanche, the diffusion speed of the carrier cluster is slow enough for the spatial distribution to be treated as Gaussian with a variance that does not change over time.
Thus, to count the carriers that reached the gain layer from a given cluster, the relation \ref{eq:time_cov} can be deduced into

\begin{equation}
    N(z_{e},t)= \frac{\sum N}{\sqrt{2\pi[\tau^2+(\sigma/v_{e})^2]}} \exp{\left[-\frac{(t-t_0)^2}{2\tau^2+2(\sigma/v_{e})^2}\right]},
\end{equation}
where $z_{e}$ refers to the bottom of the gain layer, $v_{e} = \mu E(z_{e})$ represents the drifting velocity of the carriers at that location, and $t_0 = \int_{z_i}^{z_e}dz/v$.

Consider carriers, the number of which is $\mathrm{d}N = N\mathrm{d}t$, passing through the gain layer within a small time period $\mathrm{d}t$. As a result, $M\mathrm{d}N$ ionized carrier pairs are created, which then start to drift and generate a signal. As only carriers with opposite charges can continue to drift and contribute to the signal while crossing the space charge region, the amplitude will increase by $M\mathrm{d}N q v_{e}'/d$ (where the prime denotes carriers with a different charge sign, and $d$ represents the thickness of the detector).
Therefore, the time derivative of the total gain signal, $\mathrm{d}i_{gain}/\mathrm{d}t$, is proportional to $N$, and the maximum of the derivative is also proportional to the maximum of $N$. This is because the gain signal $i_{gain}$ dominates the rising edge of the total signal $i_q$ by

\begin{equation}
    \frac{\mathrm{d}i_q}{\mathrm{d}t}\Big|_{max} = \frac{k_2 \sum N}{\sqrt{\tau^2 v_{e}^2+\sigma^2}},
    \label{gain_signal_maximum}
\end{equation}
where
$k_2 = (Mq/\sqrt{2\pi})/(v_{e}v'_{e}/d) = 2.30\times 10^{-7}\rm \frac{A \cdot \mu m}{n s}.$ Notice that for non-irradiated detectors, $k_2$ could be completely determined by the wavelength of the laser and properties of the detector, since $v_{e}$ is a function of $E_{e}$, and the value of $E_{e}$ can be estimated by $E_{e}\approx 2(V_{GD}-V_{FD})/d$ where $V_{GD}$ and $V_{FD}$ represent the full depletion voltages of the gain layer and the active area.

Considering the distribution of edge-TCT induced carriers described by Equation \ref{laser_intensity}, in Equation \ref{sigma2} or \ref{sigma2_simple}, $\sigma_0$ will be replaced by the beam waist $w(x)/2$. Additionally, it is necessary to integrate along the laser beam direction $x$ in order to evaluate the change in beam waist. Plus, to account for attenuation from Equation \ref{eq:light_absorb}, within the range $(x,x+\mathrm{d}x)$, the total number of carriers is $(E_p/h\nu)\exp{(-\alpha x)}\alpha \mathrm{d}x$. Thus,

\begin{equation}
    \frac{\mathrm{d}i_q}{\mathrm{d}t}\Big|_{max} = \int_0^{x_{max}} \frac{k_2 (E_p/h\nu)\exp{(-\alpha x)}\alpha \mathrm{d}x}{\sqrt{\tau^2 v_{e}^2+ \frac{E_e^2}{E_i^2}[\frac{w_0^2(x)}{4}  + \frac {k_1}{2}(1-\frac{E_i^2}{ E_e^2})]}}.
\end{equation}

For facilities utilized in this work, since $\tau v_{e}$ is a relatively large term (approximately 15 $\mu \rm m$) and $\frac{E(z_e)}{E(z)}$ is less than 1.1, the result should be approximately constant. Though, we could expect the diffusion profile method to be more valuable and provide more information in the case that $v_{e}$ is relatively small.

The simulated and experimental diffusion profiles of LGAD are shown in Figure \ref{fig:diffusion}, along with the corresponding theoretical values.
The reliability of the diffusion profile method for LGAD has been proven, as it accurately predicts and observes absolute results.

\section{Conclusion}

The present study has successfully established and tested the diffusion profile method for measuring the bulk field in LGAD (low-gain avalanche detector). This methodology aims to use thin detectors with a thickness approximately ten times the FWHM (full width at half maximum) of the laser beam. By calculating the temporal derivative of the waveform, valuable information about the diffusion process can be obtained, as well as its impact on the flattening of the carrier cluster. The methodology is expected to yield immediate results by using directly measurable parameters, eliminating the need for normalization.

This study effectively conducted a simulation of the edge-TCT (edge transient current technique) on non-irradiated silicon LGAD using the particle detector simulation package RASER.
The simulated waveform was used to fit the data. Subsequently, an edge-TCT scan was generated by comparing the experimental results with HPK Type 3.1 detectors (p-i-n and LGAD) to determine its reliability.
The comparison yielded successful results in terms of amplitudes, rise times, and charge collections, thereby confirming the accuracy of the simulation in edge-TCT.

A comparable simulation procedure could be implemented to examine the performance of detectors made from alternative materials, such as silicon carbide, with a specific focus on their resistance to radiation when exposed to a particular irradiation flux. Based on prior research \cite{RASER_3D}, simulation can also be used to investigate TCT tests on detectors with non-planar structures \cite{3d_TCT}.

\section*{Acknowledgement}

This work has been supported by the China Postdoctoral Science Foundation (No. 2022M710085), the State Key Laboratory of Particle Detection and Electronics (No. SKLPDE-KF-202313), and the Natural Science Foundation of Shandong Province Youth Fund (No. ZR202111120161).
We have completed this work as part of the CERN RD50 collaboration framework, with the experiment conducted at the CERN-SSD laboratory.
We would like to thank Marcos Fernández García and Michael Moll for their support during the experiment, as well as the TRACS team for their valuable cooperation and discussions.

\bibliographystyle{unsrt}
\bibliography{p4_lgad_hpk_tct.bib}

\end{document}